\documentclass[aps,prb,reprint,superscriptaddress,amsmath,amssymb,showpacs]{revtex4-1}%
\renewcommand{\AA}{\mathcal{A}}

\newcommand{\BB}{\mathcal{B}}

\newcommand{\CC}{\mathcal{C}}

\newcommand{\LL}{\mathcal{L}}

\newcommand{\UU}{\mathcal{U}}
\newcommand{\TT}{\mathcal{T}}
\newcommand{\DD}{\mathcal{D}}

\newcommand{\Tr}[1]{\text{Tr} \left\{ #1 \right\}}
\newcommand{\tr}[2]{\text{Tr}_{ #1 } \left\{ #2 \right\}}

\newcommand{\av}[1]{\left\langle #1 \right\rangle}
\newcommand{\avg}[2]{\left\langle #1 \right\rangle_{#2}}
\newcommand{\expo}[1]{\text{exp}\left( #1 \right)}

\newcommand{\tS}{\text{S}}
\newcommand{\tL}{\text{L}}
\newcommand{\tR}{\text{R}}

\newcommand{\tF}{\text{F}}
\newcommand{\tSF}{\text{SF}}
\newcommand{\tB}{\text{B}}
\newcommand{\tSB}{\text{SB}}

\newcommand{\e}{\text{e}}

\newcommand{\ph}{{\phantom{\dagger}}}
\renewcommand{\bm}[1]{\textbf{\textit{#1}}}

\newcommand{\newest}[2]{{\color{black} #1}}
\newcommand{\new}[2]{{\color{black} #1}}
 
\usepackage{braket}
\usepackage{subfigure}
\usepackage{graphicx}
\usepackage{xcolor}
\usepackage{dsfont}
\usepackage{pstricks}
\usepackage{mathrsfs}
\usepackage{bbm}
\usepackage{nicefrac}

\begin{document}

%Title of paper
\title{Nonequilibrium open quantum systems with multiple bosonic and fermionic environments: \new{A hierarchical equations of motion approach}{}}

\author{J.\ B{\"{a}}tge}
\affiliation{Institute of Physics, University of Freiburg, Hermann-Herder-Str. 3, D-79104 Freiburg, Germany}

\author{Y.\ Ke}
\affiliation{Institute of Physics, University of Freiburg, Hermann-Herder-Str. 3, D-79104 Freiburg, Germany}

  \author{C. Kaspar}
  \affiliation{Institute of Physics, University of Freiburg, Hermann-Herder-Str. 3, D-79104 Freiburg, Germany}

\author{M.\ Thoss}
\affiliation{Institute of Physics, University of Freiburg, Hermann-Herder-Str. 3, D-79104 Freiburg, Germany}
\affiliation{EUCOR Centre for Quantum Science and Quantum Computing, University of Freiburg, Hermann-Herder-Str. 3, D-79104 Freiburg, Germany}

\date{\today}

\begin{abstract}
We present a hierarchical equations of motion approach, which allows a numerically exact simulation of nonequilibrium transport in general open quantum systems involving multiple \newest{macroscopic}{} bosonic and fermionic environments. The performance of the method is demonstrated for a model of a nanosystem, which involves interacting electronic and vibrational degrees of 
freedom and is coupled to fermionic and bosonic baths. The results show the intricate interplay of electronic and vibrational degrees of freedom in this nonequilibrium transport scenario for both voltage and thermally driven transport processes. Furthermore, the use of importance criteria to 
improve the efficiency of the method is discussed. 
\end{abstract}


\maketitle
%------------------------------
%Introduction
%------------------------------

\section{Introduction}

Open quantum systems, which involve coupling to one or multiple reservoirs, are ubiquitous in many areas of physics, including nanophysics, quantum thermodynamics, quantum optics, and quantum information theory.\cite{Weiss1999,Breuer2007theory} They are also relevant for a variety of novel technological developments that involve a nanoscale system coupled to a macroscopic environment, such as quantum information devices\cite{Loss1998,Petersson2012}, nanoelectronic setups\cite{Piva2005,Song2009}, and thermoelectric generators on the scale of single molecules\cite{Dubi2011,Finch2009,Karlstroem2011}.
The theoretical description of open quantum systems requires in most cases numerical methods. In this context, several approximate methods \cite{Caroli1972, Hyldgaard1994, Galperin2006, Breuer2007theory, Mitra2004, Timm2008, Leijnse2008, Haertle2011, Ness2001, Cizek2004, CasparyToroker2007, Laakso2014, Khedri2017, Khedri2018, Liu2020} as well as numerically exact 
techniques
\new{\cite{
Huetzen2012,Weiss2013,Simine2013,Muehlbacher2008,Schiro2009,Klatt2015,Wang2009,Wilner2014,Wang2013,Wang2016,Strathearn2018,Joergensen2019,Anders2006,Jovchev2013,Werner2009,Cohen2011,Cohen2015,Eidelstein2020,Stockburger2002,Shao2004,Lacroix2005,Ke2016,Hsieh2018a,Han2020}}{\cite{Huetzen2012,Weiss2013,Simine2013,Muehlbacher2008,Schiro2009,Klatt2015,Wang2009,Wilner2014,Wang2016,Anders2006,Jovchev2013,Cohen2011,Wilner2014}} have been developed and applied. \new{Many applications of the numerically exact approaches focus on the treatment 
of either bosonic or fermionic environments. However, especially in the realm of 
quantum thermodynamics and quantum transport, the inclusion of both types of environments in the model is often necessary. So far, only a few applications of numerically exact techniques have 
been reported to scenarios which include both types of environments.\cite{Muehlbacher2008,Simine2013,Wang2009,Wang2013,Wilner2014} In this paper, we extend}{ With the exception of the path-integral approach formulated by Simine and Segal\cite{Simine2013}, to our knowledge, all applications of numerically exact approaches focus either on the treatment of bosonic or fermionic environments so far. However, especially in the realm of quantum 
thermodynamics and quantum transport the inclusion of both types of environments in the model is often necessary. To this end, we extend in this paper} the hierarchical equations of motion approach (HEOM) for efficient treatment of open quantum systems with multiple bosonic and fermionic environments.

\begin{figure}[t]
 \centering
 \includegraphics{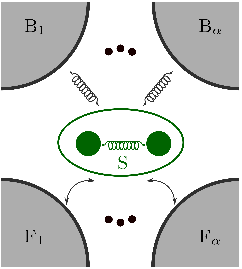}
 \caption{Sketch of an open quantum system coupled to multiple fermionic and bosonic environments.}
 \label{fig:most_general_model}
\end{figure}

The \new{HEOM method (in the context of nonequilibrium electron transport,
also called the hierarchical quantum master equation approach)}{HQME ... (HEOM)} was originally developed  by Tanimura and Kubo\cite{Tanimura1989,Tanimura1990} for the description of relaxation dynamics in open quantum systems induced by the coupling to bosonic environments. Later, it was 
extended by several groups for the investigation of charge and energy transport caused by fermionic environments.\cite{Jin2007,Jin2008,Zheng2012,Haertle2013a,Haertle2015,Huo2015,Schinabeck2016,Ye2016,Wenderoth2016,Schinabeck2018,Schinabeck2020} As summarized by Tanimura,\cite{Tanimura2020} so far studies using the HEOM approach have focused on either bosonic or fermionic \newest{reservoirs. In Ref. \onlinecite{Schinabeck2018}, the HEOM method was formulated and applied to a system coupled to an environment consisting of two fermionic reservoirs and a single bosonic mode.}{environments.}
In this work, we formulate the HEOM method for more general models, depicted in Fig. \ref{fig:most_general_model}, which include multiple bosonic and fermionic environments. This extension allows us to study 
more general scenarios such as electronic and phononic heat transport induced by bias voltages or temperature differences. 

The outline of this paper is as follows: The generalization of the model system and the HEOM formalism are introduced in Sec. \ref{sec:Methodology}.
In Sec. \ref{sec:Illustrative_applications}, we discuss the performance of the method for exemplary models, including both voltage and thermally driven charge and heat transport. 
Section \ref{sec:Conclusion} concludes with a summary.
Throughout the paper, we use units where $\hbar=1$ and $k_\tB=1$.

%------------------------------
%Methodology
%------------------------------

\section{Method}
\label{sec:Methodology}

We consider an open quantum system described by a subsystem S, which is coupled to both bosonic (B) and fermionic (F) environments.
The total Hamiltonian is given by
\begin{align}
 {H} = {H}_\tS + {H}_\tSB + {H}_\tB+ {H}_\tSF + {H}_\tF.
  \label{eq:general_Hamiltonian}
\end{align}
 The fermionic environment  ${H}_\tF$ may consist of multiple macroscopic 
noninteracting fermionic baths,
 \begin{align}
    {H}_\tF =\sum_\alpha {H}_{\tF_\alpha} = \sum_{\alpha,k} \varepsilon^\ph_{k \alpha} {c}_{k \alpha}^\dagger {c}_{k \alpha}^\ph.
    \label{eq:Hamiltonian:fermionic_bath}
  \end{align}
Here, $c^\dagger_{k \alpha} (c^\ph_{k \alpha})$ denotes the fermionic creation(annihilation) operator associated with state $k$ of bath $\alpha$ with energy $\varepsilon_{k \alpha}$. 
Nonequilibrium situations can be introduced by assigning different chemical potentials $\mu_{\tF_\alpha}$ and/or temperatures $T_{\tF_\alpha}$ to the different baths.
 Similarly, the bosonic environment $ {H}_\tB$ may contain more than one noninteracting 
 bosonic bath,
   \begin{align}
    {H}_\tB = \sum_\alpha {H}_{\tB_\alpha} = \sum_{\alpha, j } \omega^\ph_{j \alpha} {b}_{j \alpha}^\dagger {b}_{j \alpha}^\ph,
    \label{eq:Hamiltonian:bosonic_bath}
  \end{align}
 where $b^\dagger_{j \alpha}(b^\ph_{j \alpha})$ denotes the bosonic creation(annihilation) operator associated with state $j$ of bath $\alpha$ with 
frequency $\omega_{j \alpha}$. 
 Nonequilibrium situations can be achieved by choosing different temperatures $T_{\tB_\alpha}$ for different bosonic baths. For simplicity, we assume vanishing chemical potentials $\mu_{\tB_\alpha}=0$, which holds true for phonons and photons.
  
 For the validity of the subsequent derivation of the HEOM, we restrict ourselves to system-environment coupling terms which are linear in the bath creation and annihilation operators,
 \begin{subequations}
  \begin{alignat}{2}
    {H}_\tSF  =&  \sum_{\alpha,k}   \nu^\ph_{k \alpha} c_{k \alpha}^\dagger W + \text{H.c.},
    \label{eq:Hamiltonian:fermionic_coupling}
    \\
    {H}_\tSB  =& {V} \sum_{\alpha,j} \xi^\ph_{j \alpha} \left(  {b}_{j \alpha}^\dagger + {b}_{j \alpha}^\ph\right).
    \label{eq:Hamiltonian:bosonic_coupling}
  \end{alignat}
 \end{subequations}  
 Both $V$ and $W$ are operators acting only on the subsystem degrees of freedom (DOFs).
 They are not restricted in their action on the bosonic DOFs; however, $W$ has to act as a generalized annihilation operator on the fermionic DOFs and overall reduce one fermionic occupation number by one.
 In contrast, $V$ must not change the fermionic occupation numbers and may involve sums of even numbers of fermionic annihilation and creation operators.
 The influence of the bosonic and fermionic environment on the system dynamics is encoded in the so-called spectral density functions,
 \begin{subequations}
  \begin{align}
    J_{\tB_\alpha}(\omega) =& \pi \sum\limits_{j} |\xi^\ph_{j \alpha}|^2 \delta(\omega-\omega^\ph_{j \alpha}) ,
    \label{eq:bosonic_spectral_density_general_definition}
    \\
    J_{\tF_\alpha}(\varepsilon) =& 2 \pi \sum\limits_{k} |\nu^\ph_{k \alpha}|^2 \delta(\varepsilon-\varepsilon^\ph_{k \alpha})  .
    \label{eq:fermionic_spectral_density_general_definition}
  \end{align}
 \end{subequations}

 With these general specifications of the open quantum system, we present 
the general idea of the derivation of the HEOM, which is outlined in more 
detail in Refs. \onlinecite{Tanimura1989,Jin2008}.
 The central quantity of the approach is the reduced density matrix $\rho(t)$ of the subsystem, defined by the trace over the DOFs of the environment of the total density matrix $\varrho(t)$,
 \begin{align}
  \rho(t) = \tr{\tB+\tF}{\varrho(t)}.
 \end{align}
 Employing a bath-interaction picture, we  avoid  the  direct  appearance 
 of bath related dynamical phases in the equation of motion of the reduced density matrix.
 In this picture, the time-dependence of the reduced density matrix can be written as
 \begin{align}
  \rho(t) = \tr{\tB+\tF}{\UU(t,0) \varrho(0) \UU^\dagger(t,0)},
 \end{align} 
 with the time-evolution operator
 \begin{align}
  \UU(t,t_0) = \TT\left(\e^{-i\int_{t_0}^t d \tau H_\tSB(\tau) + H_\tSF(\tau) + H_\tS} \right).
 \end{align}
 Here, $\TT$ denotes the time-ordering operator and the time-dependence of $H_\tSB$/$H_\tSF$ originates from the chosen bath-interaction picture.

 Exploiting the Gaussian properties of the noninteracting \new{macroscopic}{} environments, a 
formally exact path-integral for the reduced density matrix of the subsystem involving a Feynman-Vernon influence functional can be derived.\cite{Feynman1951,Feynman1963,Tanimura1989,Jin2008} 

 As a consequence of noninteracting Gaussian baths, all information about 
system-bath couplings is encoded in the time correlation functions of the 
free baths
 \begin{subequations}  
 \begin{align}
    C_{\tB_\alpha}(t-\tau)=&\sum\limits_{j} |\xi^\ph_{j \alpha}|^2 \avg{   {b}_{j \alpha}^\dagger (t) {b}_{j \alpha}^\ph(\tau)  +  {b}_{j \alpha}^\ph(t) {b}_{j \alpha}^\dagger(\tau) }{\tB_\alpha} ,
    \\
    C^s_{\tF_\alpha}(t-\tau)=&\sum\limits_{k } |\nu_{k \alpha}|^2 \avg{ 
 c^{s}_{k \alpha}(t)  c^{\bar{s}}_{k \alpha}(\tau) }{\tF_\alpha} .
 \end{align}
 \end{subequations}  
 Here, $\avg{\,\cdot\,}{X_\alpha} = \tr{X_\alpha}{\,\cdot \;\rho_{X_\alpha}}$ denotes the expectation value with respect to the corresponding free bath 
 \new{ \begin{equation}\rho_{X_\alpha} = \frac{\text{exp}\left(-\beta_{X_\alpha} \left( H_{X_\alpha}-\mu_{X_\alpha}N_{X_\alpha}\right)\right)}{\tr{X_\alpha}{\text{exp}\left(-\beta_{X_\alpha} \left( H_{X_\alpha}-\mu_{X_\alpha}N_{X_\alpha}\right)\right)}},\end{equation}}{$\rho_{X_\alpha} = \frac{\text{exp}\left(-\beta_{X_\alpha} \left( H_{X_\alpha}-\mu_{X_\alpha}N_{X_\alpha}\right)\right)}{\tr{X_\alpha}{\text{exp}\left(-\beta_{X_\alpha} \left( H_{X_\alpha}-\mu_{X_\alpha}N_{X_\alpha}\right)\right)}}$}
 and $s\in\{-,+\}$ has been introduced with $c^{-}_{k \alpha}=c^{\ph}_{k \alpha}$ and $c^{+}_{k \alpha}=c^{\dagger}_{k \alpha}$. Furthermore, $N_{X_\alpha}$ represents the particle number operator of the bosonic or fermionc particles and $\beta_{X_\alpha}$ describes the inverse temperature $1/T_{X_\alpha}$  in the respective bath $X_\alpha$.
The time correlation functions are related to the bath spectral densities 
(see Eqs. \eqref{eq:bosonic_spectral_density_general_definition} and \eqref{eq:fermionic_spectral_density_general_definition} ) 
via 
\begin{subequations}  
 \begin{align}
  C_{\tB_\alpha}(t) = & \int_0^\infty d\omega\frac{J_{\tB_\alpha}(\omega)}{\pi} \nonumber \\
      &\times \left[\text{coth}\left(\frac{\beta_{\tB_\alpha} \omega}{2}\right)\text{cos}(\omega t) - i \text{ sin}(\omega t)\right],
      \\
  C^s_{\tF_\alpha}(t) = & \int_{-\infty}^{\infty} d\varepsilon \frac{J_{\tF_\alpha}(\varepsilon)}{2\pi}\frac{ \e^{s i \varepsilon t}}{\e^{s\beta_{\tF_\alpha} (\varepsilon-\mu_{\tF_\alpha})}+1} .
 \end{align}
\end{subequations}  
Within the HEOM approach, the correlation functions are described by sums 
over exponentials,
 \begin{subequations}  
    \begin{align}
    C_{\tB_\alpha}(t) = \Lambda_{\tB_\alpha} \sum_{p} \eta_{p,\alpha} \e^{-\gamma_{p, \alpha} t},\\
    C^s_{\tF_\alpha}(t) = \Gamma_{\tF_\alpha} \sum_{q} \eta_{q,\alpha,s} \e^{-\gamma_{q, \alpha ,s} t},
    \end{align}
 \end{subequations}  
where common  strategies to determine the weights $\eta$ and decay rates $\gamma$ include the  Matsubara\cite{Tanimura1989,Tanimura1990,Ishizaki2005,Tanimura2006,Jin2008,Moix2013,Erpenbeck2018} and  Pad\'e\cite{Hu2010,Hu2011,Kato2015,Erpenbeck2019,Schinabeck2020}  decompositions.
Furthermore, $\Lambda_{\tB_\alpha}$ and $\Gamma_{\tF_\alpha}$ are coupling strengths of the respective baths and are defined by the respective spectral density. The exponentially decaying time correlation functions ensure a closed set of equations of motion.\cite{Tanimura2020}

Under the constraints on the Hamiltonian stated above, the influence functional factorizes into two parts which originate from the bosonic and fermionic environments separately. \cite{Zheng2012,Schinabeck2018} The separate influence functionals can be obtained as shown in detail in previous works.\cite{Tanimura1989,Tanimura1990,Jin2008,Schinabeck2018} 
The hierarchical equations of motion are constructed by consecutive time derivatives of the influence functional. The combination of the hierarchies originating from the different environments leads to\cite{Schinabeck2018}
\begin{alignat}{2}
   \frac{\partial}{\partial t} \rho^{(m|n)}_{\bm{g}|\bm{h}} =& -\left( i \LL_\tS + \sum_{l=1}^m \gamma_{g_l} + \sum_{l=1}^n \gamma_{h_l} \right) \rho^{(m|n)}_{\bm{g}|\bm{h}}
   \nonumber \\
   &- \sum_{h_x} \AA_{h_x} \rho^{(m|n+1)}_{\bm{g}|\bm{h}^+_x} -\sum_{l=1}^n (-1)^l \CC_{ h_l}  \rho^{(m|n-1)}_{\bm{g}|\bm{h}^-_l}
   \nonumber \\
   &+ \sum_{g_x} \BB_{g_x} \rho^{(m+1|n)}_{\bm{g}^+_x|\bm{h}} +\sum_{l=1}^m \DD_{ g_l}  \rho^{(m-1|n)}_{\bm{g}^-_l|\bm{h}},
   \label{eq:general_HEOM}
\end{alignat}
with the multi-indices $g=(p,\alpha)$ and $h=(q,\alpha,s)$, the notation for the multi-index vectors ${\bm{v}} = {v_1 {\cdot}{\cdot}{\cdot}v_p}$, ${\bm{v}^+_x}= {v_1 {\cdot}{\cdot}{\cdot}v_p v_x}$, and ${\bm{v}^-_l} = {v_1 {\cdot}{\cdot}{\cdot}v_{l-1}v_{l+1}{\cdot}{\cdot}{\cdot}v_p}$,\footnote{For the bosonic hierarchy the index order is not relevant. 
In contrast the index order is important for the fermionic hierarchy.} and $\LL_\tS O = [H_\tS,O]$.
To facilitate a unified treatment of the bosonic and fermionic hierarchy, the index management for the bosonic part of the hierarchy differs from the convention in other publications.\cite{Kato2015,Kato2016,Song2017}
Here, $\rho^{(0|0)}$ represents  the reduced  density operator of the subsystem, and the higher-tier auxiliary density operators (ADOs) $\rho^{(m|n)}_{\bm{g}|\bm{h}}$ encode the influence of the environment on the subsystem dynamics. Tier by tier the ADOs introduce the effect of higher order 
system-bath correlations.  These ADOs also include the information on time-local transport  observables like energy and particle currents.\cite{Jin2008,Haertle2013a,Kato2015,Kato2016}  Employing the generating functional 
approach\cite{Schwinger1961,Kato2016,Song2017} such observables are related to compositions of ADOs. 

The operators $\AA_{h_x}$ and $\CC_{h_l}$ connect the $n$th-fermionic-tier ADO to the ($n{+}1$)th- and $(n{-}1)$th-fermionic-tier ADOs, respectively, and the operators $\BB_{g_x}$ and $\DD_{g_l}$ connect the $m$th-bosonic-tier ADO to the ($m{+}1$)th- and $(m{-}1)$th-bosonic-tier ADOs via
 \begin{subequations}  
\begin{align}
 \AA_{h_x} \rho^{(m|n)}_{\bm{g}|\bm{h}}=&  \Gamma \left( W^{s_{h_x}}  \rho^{(m|n)}_{\bm{g}|\bm{h}} + (-1)^{(n)} \rho^{(m|n)}_{\bm{g}|\bm{h}} W^{s_{h_x}} \right),\\
 \BB_{g_x} \rho^{(m|n)}_{\bm{g}|\bm{h}}=&  \Lambda  \left[ {V} , \rho^{(m|n)}_{\bm{g}|\bm{h}}\right],\\
 \CC_{h_l} \rho^{(m|n)}_{\bm{g}|\bm{h}}=&\, (-1)^{n} \eta_{h_l}   {W^{\bar{s}_{h_l}} }  \rho^{(m|n)}_{\bm{g}|\bm{h}} - \eta^*_{ \bar{h}_l}  \rho^{(m|n)}_{\bm{g}|\bm{h}} {W}^{\bar{s}_{h_l}}, \\
 \DD_{g_l} \rho^{(m|n)}_{\bm{g}|\bm{h}}=&\, \eta_{g_l}   {V}  \rho^{(m|n)}_{\bm{g}|\bm{h}} - \eta^*_{ g_l}  \rho^{(m|n)}_{\bm{g}|\bm{h}} {V}, 
 \label{eq:general_HEOM_upbuilding_operators}
\end{align}
 \end{subequations}  
leading to a twofold hierarchy of equations of motion. Generalizing these operators for linear superpositions of the system-bath couplings defined in Eqs. (\ref{eq:Hamiltonian:bosonic_coupling}) and (\ref{eq:Hamiltonian:fermionic_coupling}), the superindices $g$ and $h$, the operators $V$ and $W$ and the coupling strengths $\Lambda$ and $\Gamma$ acquire an 
additional index.  

For applications, only a finite number of poles characterizing the baths can be taken into account. By employing additional resummation schemes this limitation can be overcome.\cite{Erpenbeck2018}  Here, the hierarchy needs to be truncated in a suitable manner. For this purpose previous works have introduced an importance criterion for pure fermionic environments, which estimates the importance of an ADO to the dynamics of the system.\cite{Shi2009,Haertle2013a,Wenderoth2016} If the importance $\mathcal{I}$ 
assigned to an ADO is below a given threshold value $\theta$, %???maybe a different symbol???
it is neglected.  
 This reduces the numerical complexity to a manageable level and allows for a systematic convergence. For details of the convergence properties of 
the HEOM method, we refer to  Refs. \onlinecite{Popescu2015,Haertle2015,Dunn2019,Rahman2019}.  Here, we generalize the criterion to the combined hierarchy and assign to each operator $\rho^{(m|n)}_{\bm{g}|\bm{h}}$ the importance value
 \begin{align}
	 \mathcal{I} \left( \rho^{(m|n)}_{\bm{g}|\bm{h}}\right) =&  \left|\prod\limits_{l=1}^{n }\frac{\Gamma}{\sum\limits_{a\in\{1..{l}\}}\hspace{-.2cm} \text{Re}\left[\gamma_{h_{a}}\right]}   \frac{\eta_{h_{l}}}{\text{Re}\left[\gamma_{h_{l}}\right]} \right| \nonumber \\
	 &\times	 \left| \prod\limits_{l=1}^{m }\frac{\Lambda}{\sum\limits_{a\in\{1..{l}\}}\hspace{-.2cm} \text{Re}\left[\gamma_{g_{a}}\right]}  \frac{\eta_{g_{l}}}{\text{Re}\left[\gamma_{g_{l}}\right]} \right|.
	 \label{eq:importance_estimate}
\end{align}
 The importance of an ADO is weighted according to its influence within the fermionic and bosonic hierarchies simultaneously. 
 Motivated by the factorized structure of the HEOM, we add a factor considering the bosonic part of the twofold hierarchy to the importance estimate for the ADOs in the fermionic hierarchy presented by H\"artle \textit{et al.}, which stems from the HEOM in the stationary condition.\cite{Haertle2013a}  This factor is constructed similarly to that for the fermionic environment. 
 It assesses the importance due to the respective order in the bosonic part of the hierarchy, and it reflects the relative importance of an ADO within one hierarchical order for the stationary state.

%------------------------------
%Results
%------------------------------
%------

\section{Illustrative applications}
\label{sec:Illustrative_applications}

 In this section, we illustrate the performance of the HEOM approach for the combination of bosonic and fermionic environments. To this end, we consider a basic model for the subsystem involving interacting electronic and vibrational degrees of freedom. The results of the numerically exact HEOM method are compared to the well-established Born-Markov master equation approach (BMME). Explicit formulas and detailed derivations for the BMME can be found in Refs.\ \onlinecite{Breuer2007theory,Mitra2004,Haupt2006,Haertle2011}. The BMME is perturbative in the system-environment coupling up to the second, i.e., lowest nonvanishing, order and assumes a separation of relaxation timescales in the bath and the system. 
 \new{The comparison allows a validation of the approximate BMME method and underlines the necessity to use numerically exact methods such as HEOM for stronger-coupling regimes.}{}
 
\subsection{Model}
\label{sec:IA:Model}

\begin{figure}[t]
  \centering
  \includegraphics[width=0.3\textwidth]{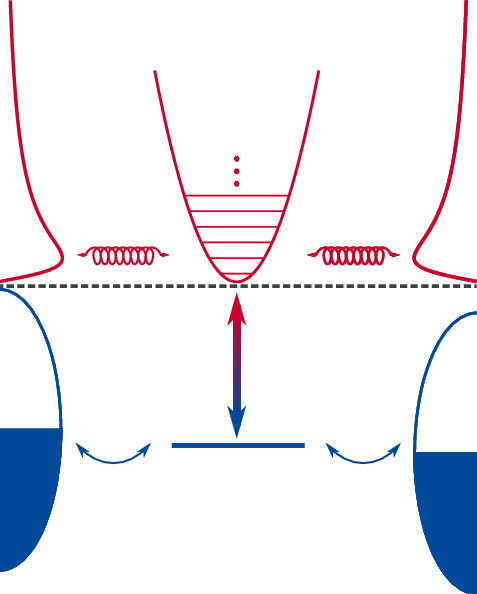}
  \caption{Sketch of the model consisting of one electronic state and a single vibrational mode coupled to two bosonic and two fermionic baths, describing vibrational heat baths and electronic leads, respectively.}
  \label{fig:generic_model}
 \end{figure}
 As sketched in Fig. \ref{fig:generic_model}, we consider a model system that consists of a single electronic state and a single vibrational mode, 
described by the Hamiltonian \cite{Koch2005,Haertle2018,Schinabeck2018,Schinabeck2020}
 \begin{align}
  H_\tS = \Omega a^\dagger a + \varepsilon_0 d^\dagger d + \lambda (a^\dagger + a ) d^\dagger d  .
  \label{eq:Hamiltonian:specific_system_Anderson_Holstein}
 \end{align}
 Here, $a^\dagger$ and $a$ are the creation and annihilation operators of 
the vibrational mode with frequency $\Omega$ and $d^\dagger$ and $d$ are the creation and annihilation operators of the electronic state with energy $\varepsilon_0$. The electronic state is coupled to the vibrational mode via the coupling constant $\lambda$. The couplings to the respective environments have the forms
 \begin{subequations}  
  \begin{alignat}{2}
    {H}_\tSF  =&  \sum_{\alpha,k} \left( \nu^\ph_{k \alpha} c_{k \alpha}^\dagger d + \text{H.c.} \right),
    \label{eq:Hamiltonian:specific_fermionic_coupling}
    \\
    {H}_\tSB  =& (a^\dagger +a )\sum_{\alpha,j} \xi^\ph_{j \alpha} \left(  {b}_{j \alpha}^\dagger + {b}_{j \alpha}^\ph\right),
    \label{eq:Hamiltonian:specific_bosonic_coupling}
  \end{alignat}
 \end{subequations}  
 and are specified by the spectral densities
 \begin{subequations}  
 \begin{align}
    J_{\tB_\alpha}(\omega)% =& \pi \sum\limits_{j\in \tB_i} |\xi^\ph_j|^2 \delta(\omega-\omega^\ph_j) \\
    =& \Lambda_{\tB_\alpha} \frac{\omega}{\Omega} \frac{ \omega^2_{c \alpha}}{\omega_{c \alpha}^2+\omega^2}
    \label{eq:bosonic_spectral_density_Drude_definition}
    \\
    J_{\tF_\alpha}(\varepsilon)% =& 2 \pi \sum\limits_{k \in \tF_i} |\nu^\ph_k|^2 \delta(\varepsilon-\varepsilon^\ph_k) \\
    =& \Gamma_{\tF_\alpha} \frac{D_{\alpha}^2}{D_{\alpha}^2+(\varepsilon-\mu_{\tF_\alpha})^2} .
    \label{eq:fermionic_spectral_density_Lorentzian_band_definition}
 \end{align}
 \end{subequations}  
 Here, we include the system frequency $\Omega$ in the definition of the bosonic bath spectral density for two purposes. First, the behavior of the spectral density in the limit $\omega \to 0$ does not depend on the cutoff frequency $\omega_{c \alpha}$. Second, the coupling strength $\Lambda_{\tB_\alpha}$ has the dimension of an energy and defines a decay time.\cite{Breuer2007theory} %compare Breuer & Petruccione p. 481 
 The Lorentzian suppression of high-energy contributions by the cutoff frequency and the bandwidth $D_{\alpha}$ is beneficial for the closed derivation of the HEOM
 %is beneficial for the numerical evaluation of the HEOM
 and allows for a straightforward use of the  Pad\'e decomposition scheme 
to approximate the correlation functions.\cite{Xie2012,Kato2015}
 Throughout this work, we assume symmetric configurations, i.e., $\Gamma_\tL=\Gamma_\tR=\frac{\Gamma}{2}$, and $\Lambda_\tL=\Lambda_\tR=\frac{\Lambda}{2}$.
 
 It is noted that this model has been studied extensively without bosonic 
baths using both approximate and numerically exact approaches.\cite{Koch2004,Mitra2004,Koch2005,Haertle2009,Haertle2011,Haertle2018,Schinabeck2018,Schinabeck2020}
 For the explanation of the transport phenomena, the system part of the Hamiltonian is often diagonalized by the small polaron transformation via $\overline{H} = S H S^\dagger$, with $S=\expo{\frac{\lambda}{\Omega}d^\dagger d (a-a^\dagger)}$. 
 Thereby, the charging and decharging processes of the system induced by the fermionic environment are accompanied by de-/excitations of the vibrational mode with their transition amplitudes given by the so-called Franck-Condon matrix elements\cite{Koch2004,Mitra2004,Koch2005,Koch2006}, and the electronic-vibrational interaction term in the system transforms to a shift of the electronic state energy $\overline{\varepsilon}_0=\varepsilon_0 - \frac{\lambda^2}{\Omega}$.
 Applying the small polaron transformation also affects the coupling term 
to the bosonic environment. Accordingly, the shift of the electronic state energy and the Franck-Condon elements are no longer exact but perturbative up to the lowest nonvanishing order in the system-environment coupling. A non-perturbative generalization of the transformation leads to a non-linear system-environment coupling.\cite{Haupt2006}
 Therefore, we do not use such a transformation for the combined fermionic and bosonic HEOM considered here.
 
 Subsequent to the introduction of the observables of interest, we show results for two parameter regimes specified in more detail in Table \ref{tab:definition_parameter_sets}. Parameter set A is chosen such that the perturbative treatment within the BMME is expected to be valid. 
 In contrast, the system-environment coupling is enhanced, and the cutoff frequency in the bosonic bath $\omega_{c \alpha}$ is reduced in parameter set B. Both changes weaken the validity of the perturbative treatment of the model system by the BMME. 
  \begin{table}[t]
  \centering
  \begin{tabular}{ccccccccc}
   \hline \hline
   Set &     $\Gamma$  & $D_{\alpha}$&  $\Lambda$ &  $\omega_{c \alpha}$ & $T$ & $\Omega$    & $\overline{\varepsilon}_0$  \\ \hline 
   A  &   0.005 &   30   & 0.005    &   0.2     & 0.025    & 0.2   & 0.2 \\ 
   B  &   0.05  &   30   & 0.0425   &   0.05    & 0.025    & 0.2   & 0.2 \\
   \hline \hline
  \end{tabular}
 \caption{Model parameters used in the calculations. All parameters are given in $e$V.}
 \label{tab:definition_parameter_sets}
 \end{table}

 \subsection{Observables}
 \label{sec:IA:Observables}
 System observables, such as the population of the electronic level and the vibrational excitation, can be directly evaluated from the reduced density matrix of the system. For environment-related observables, we employ the generating functional approach\cite{Schwinger1961,Kato2016,Song2017} to find expressions in terms of elements of the hierarchy.
 For the charge current from lead $\tF_\alpha$ into the system, we obtain\cite{Jin2008,Haertle2013a}   
 \begin{alignat}{2}
 I_{\tF_\alpha} &= - e \av{\frac{d N_{\tF_\alpha} }{d t}} &&=  e\,\Gamma_{F_\alpha} \sum_{h_\alpha} s_h \Tr{ d^{\bar{s}_{h_\alpha}} \rho^{(0|1)}_{\ph|{h_\alpha}} }.
 \end{alignat}
 The electronic (vibrational) heat current from lead $\tF_\alpha$ (bath $\tB_\alpha$) into the system is given by\cite{Kato2015,Kato2016} 
 \begin{subequations}  
 \begin{align}
 J_{\tF_\alpha} &= -\av{\frac{d H_{\tF_\alpha}}{d t}} + \mu_{\tF_\alpha} \av{\frac{d N_{\tF_\alpha} }{d t}} \nonumber \\
           &= -i\Gamma_{\tF_\alpha} \sum_{h_\alpha} \gamma_{h_\alpha} \Tr{ d^{\bar{s}_{h_\alpha}} \rho^{(0|1)}_{\ph|h_\alpha}  } -\frac{\mu_{\tF_\alpha}}{e}  I_{\tF_\alpha}, \label{eq:fermionic_heat_current} \\
 J_{\tB_\alpha} &= -\av{\frac{d H_{\tB_\alpha}}{d t}}  \nonumber \\
           &=  i\Lambda_{\tB_\alpha} \sum_{g_\alpha} \gamma_{g_\alpha} \Tr{(a+a^\dagger) \rho^{(1|0)}_{g_\alpha|}} \nonumber \\
           &\quad+ \text{Im}\left\{C_{\tB_\alpha}(0) \right\}\Tr{(a+a^\dagger)^2 \rho^{(0|0)} }.
           \label{eq:bosonic_heat_current}
 \end{align}
 \end{subequations}  
It should be noted that there exist different definitions of the heat current.  For example, the expression $J_{\tB_{\alpha}}=-\av{\frac{d H_{\tB_\alpha}}{d t}}- \av{\frac{d H_{\tSB_\alpha}}{dt}}$ 
was used in Refs. \onlinecite{VELIZHANIN2008,Song2017} for bosonic heat currents. \new{ Furthermore, in Ref. \onlinecite{Esposito2015}, the transient and thermodynamic properties of different heat current definitions in a fermionic transport setup were discussed.}{}
In the steady state, which is considered in this work, these different definitions are equivalent to Eqs. (\ref{eq:fermionic_heat_current}) and (\ref{eq:bosonic_heat_current}), because the term $\av{\frac{d H_{\tSB_\alpha}}{dt}}$ vanishes.

 It can be shown that  expectation values of observables, which are quadratic in system and/or environment annihilation or creation operators, are 
exact for a noninteracting system ($\lambda=0$) if all contributions of fermionic and bosonic ADOs up to second tier are taken into account in the calculation. This statement includes the charge and heat currents and 
is a generalization of the statement for fermionic model systems.\cite{Jin2008,Haertle2013a}
 For the interacting system considered here, the tier of ADOs required for convergence is checked in test calculations.
 
 \subsection{Voltage-driven charge and heat transport}
 \label{sec:IA:Bias_voltage}
 
First, we consider charge and heat transport driven by an external bias voltage.
We start by recapitulating the well known charge current-voltage characteristics of the model system\cite{Mitra2004,Koch2005,Haertle2011} and briefly discuss the changes induced by the bosonic environment.\cite{Haupt2006} To this end, we expose the model system to a symmetric external  bias voltage $\Phi$, i.e., $\mu_\text{L}=-\mu_\text{R}=\frac{e\Phi}{2}$. As 
we do not apply a temperature difference and are not considering any other asymmetries of the contacts, the effect of the bosonic baths can be described by a single bath with the coupling strengths $\Lambda = \Lambda_\text{L}+\Lambda_\text{R}$ and the environmental temperature $T$.
 
 \begin{figure}[t]
  \centering
  \footnotesize
   \includegraphics{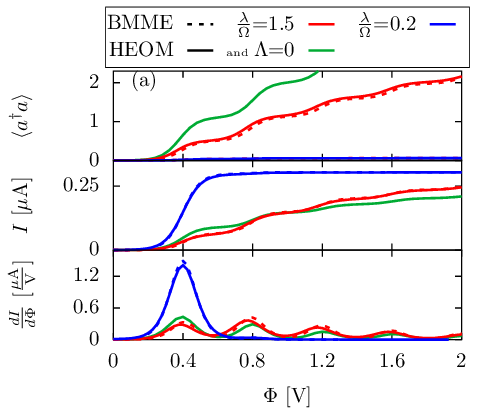}
   \includegraphics{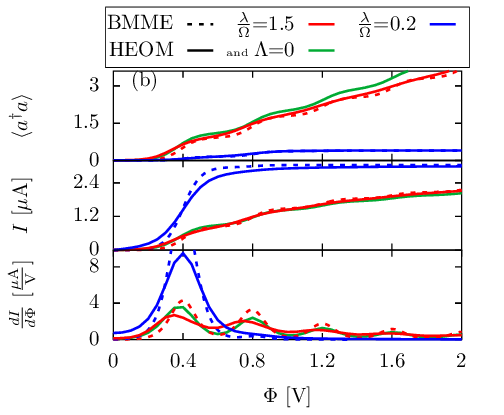}
  \caption{
    Vibrational excitation $\braket{a^\dagger a}$, charge current $I$, and differential conductance $\frac{d I}{d \Phi}$ as a function of bias voltage calculated for two different electronic-vibrational interaction strengths $\lambda$. Numerically exact HEOM results (solid lines) are compared to perturbative BMME results (dashed lines). In (a) and (b), the parameters are chosen according to parameter sets A and B, respectively. 
    The green solid line shows the HEOM results without a bosonic environment ($\Lambda=0$) for $\frac{\lambda}{\Omega}=1.5$.
  }
  \label{fig:charge_current_and_conductance_bias_voltage_induced}
 \end{figure}
Figure\ \ref{fig:charge_current_and_conductance_bias_voltage_induced} presents the vibrational excitation $\braket{a^\dagger a}$, the charge current $I$, and the differential conductance $\frac{d I}{d \Phi}$ as a function of bias voltage  $\Phi$ for two different electronic-vibrational interaction strengths, $\frac{\lambda}{\Omega}{=}1.5$ (red lines) and $\frac{\lambda}{\Omega}{=}0.2$ (blue lines). Figures \ref{fig:charge_current_and_conductance_bias_voltage_induced}(a) and \ref{fig:charge_current_and_conductance_bias_voltage_induced}(b) correspond to the results for parameter sets A and B in Table \ref{tab:definition_parameter_sets}, respectively.  Furthermore, the effect of additional vibrational relaxation is illustrated by the comparison to results without a bosonic environment ($\Lambda=0$).
The typical Franck-Condon step structure in the vibrational excitation- and the charge-current-voltage characteristics\cite{Mitra2004,Koch2005,Haertle2011} is slightly visible for weak electron-vibrational interaction ($\frac{\lambda}{\Omega}=0.2$) and becomes more pronounced for stronger interaction strengths ($\frac{\lambda}{\Omega}=1.5$). 
 The peaks in the differential conductance-voltage characteristics are located at $e\Phi \approx 2(\overline{\varepsilon}_0 + n\Omega)$, with $n\in	\mathbb{N}$. At these bias voltages, resonant transport processes become 
active.
The reduction in step heights for larger bias voltages and the corresponding saturation of the charge current are related to the structure of the Franck-Condon matrix.\cite{Haertle2011}
 
Without relaxation of the vibrational mode by a bosonic bath ($\Lambda=0$), another prominent phenomenon is the vibrational instability, which describes very high excitations of the vibrational mode for large bias voltages and, especially, for weak electronic-vibrational interactions.\cite{Koch2006b,Haertle2009} As pointed out before,\cite{Haertle2011,Haertle2011b} this effect is based on the dominance of  electron-hole pair creation processes for the cooling mechanism.
Introducing the additional vibrational relaxation by an environment, this 
dominance is reduced with increasing  coupling strength.\cite{Haupt2006,Haertle2018} As a result, the vibrational excitation for weak electronic-vibrational interaction is lower than for stronger interactions in the regime of large bias voltages.
Furthermore, the vibrational relaxation reduces the population of highly vibrational excited states of the system.
This affects the charge transport processes for strong electronic-vibrational interaction in two different ways. For small bias voltages, processes starting with an excited vibration are less likely, and the charge current is accordingly lowered.
On the contrary, transitions with a high energy transfer into the vibration are favored by the electronic-vibrational interaction for higher bias voltages. Their \new{probability}{likelihood} is enhanced by the vibrational environment, and consequently, the charge current is increased.
In addition, it is known that the damping of a bare harmonic oscillator by a bosonic bath leads to a reduction of the harmonic oscillator frequency.\cite{Grabert1988,Weiss1999} The effect of this frequency renormalization is reflected in the shift of the peak positions in the differential conduction to lower bias voltages with the introduction of the bosonic bath.
 
For weak system-environment coupling, the Born-Markov approximation is expected to be valid. This is confirmed by the agreement of the BMME and the HEOM results in this parameter regime (see Fig.\ \ref{fig:charge_current_and_conductance_bias_voltage_induced} (a)). In contrast, stronger coupling strengths increase the importance of cotunneling and other higher-order processes, which result in a broadening of the electronic energy level and, for interacting systems,  in an additional shift of the characteristic energies of transport processes (see Fig.\ \ref{fig:charge_current_and_conductance_bias_voltage_induced} (b)).  As the BMME allows only for 
sequential tunneling processes, it neglects these effects, and deviations to the numerically exact HEOM become more pronounced for stronger coupling strengths.

 \begin{figure}[t]
  \centering
  \footnotesize
   \includegraphics{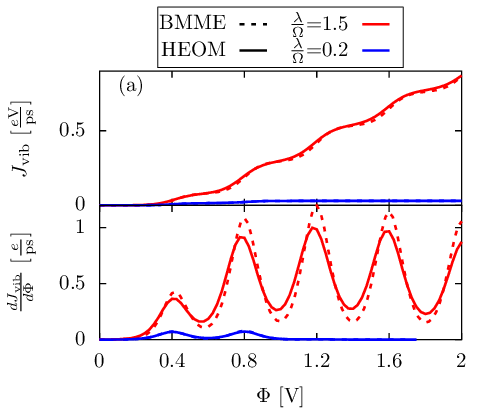}
   \includegraphics{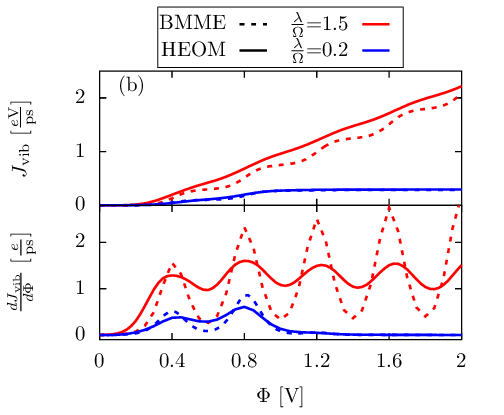}
  \caption{
    Heat current $J_\text{vib}$ from the electronic into the bosonic environment as a function of bias voltage and its differential behavior.
    In (a) and (b), the parameters are chosen according to parameter sets A and B, respectively. 
  }
  \label{fig:heat_dissipation_and_differential_dissipation_bias_voltage_induced}
 \end{figure}
Besides differences in the charge current, the bosonic environment also induces a heat current from the electronic into the vibrational degrees of 
freedom. 
Figure\ \ref{fig:heat_dissipation_and_differential_dissipation_bias_voltage_induced} depicts the heat current $J_\text{vib}=-\sum_\alpha J_{\tB_\alpha}$ from the electronic environment into the vibrational environment 
and its differential behavior as a function of bias voltage. We observe a 
steplike increase of the vibrational excitation and the heat current with bias voltage, which is in accordance with the expected excitation of the vibrational mode and its damping by the vibrational environment.
Analogous to the charge current, the steps in the heat current are connected to the onset of resonant emission processes that involve more and more vibrational quanta.
In addition, a  decrease  in  the  relative  step  heights with decreasing electronic-vibrational interaction is observed, which is caused by the suppression of processes that  involve  multiple  vibrational  quanta. However, the step heights remain almost constant with increasing bias voltage. The latter observation is a result of the interplay of the reduction of the Franck-Condon matrix elements for processes with a higher energy transfer into the vibrational degree of freedom\cite{Koch2005} and the increase of the transferred energy by these processes.

 Increasing the non-Markovianity of the environmental damping by reducing 
the cutoff frequency $\omega_c$ of the bosonic bath and increasing the coupling strengths to both environments, the BMME results deviate significantly from the numerically exact HEOM results (see Fig.\ \ref{fig:heat_dissipation_and_differential_dissipation_bias_voltage_induced} (b)). Again, 
we observe that the higher-order processes included in the HEOM lead to a 
broadening in the arising step structure and a shift of the step positions.

 \subsection{Thermally driven charge and heat transport}
 \label{sec:IA:Temperature_difference}

Another important type of transport process in nanostructures is charge 
and heat transport induced by a temperature difference of the reservoirs.
For purely electronic transport, such processes result in thermoelectric phenomena.
Here, we apply temperature differences $\Delta T$ by heating the left fermionic and bosonic bath $T_\text{L} = T + \Delta T$ without introducing an external bias voltage. The temperature of the right reservoirs remains constant.

  \begin{figure}[b]
  \centering
  \footnotesize
   \includegraphics{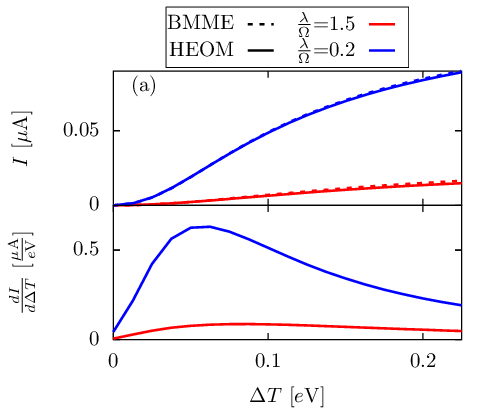}
   \includegraphics{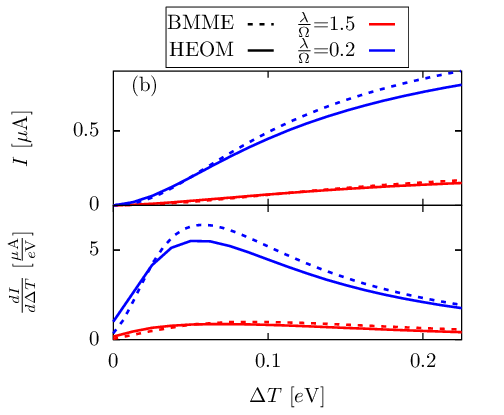}
  \caption{Charge current $I$ and its differential behavior with respect to the applied temperature difference. In (a) and (b), the parameters are chosen according to parameter sets A and B, respectively. 
}
  \label{fig:charge_current_and_conductance_temperature_difference_induced}
 \end{figure}
Figure\ \ref{fig:charge_current_and_conductance_temperature_difference_induced} shows the charge current and its differential behavior as a function of the applied temperature difference. 
Overall, the thermally driven charge current does not exhibit such pronounced features like the bias voltage induced charge current because the temperature is a smoother driving force.
Nevertheless, we observe a suppression of the charge current for stronger 
electronic-vibrational interactions, which occurs due to the Franck-Condon blockade of transport channels. 
For weak electronic-vibrational interaction, the thermally driven charge transport is mainly carried by elastic processes.
The peak in the differential conductance can be understood by the difference between the Fermi-distribution functions evaluated at the energy of the polaron shifted electronic state $\overline{\varepsilon}_0$.
In contrast, many different transport channels are contributing to the charge current in the strong electronic-vibrational interaction case, as discussed for the bias voltage driven transport in the last section. As these processes have different and higher threshold energies, their smooth onset induced by the temperature difference reduces the peak in the differential behavior of the charge current.

For weak coupling, the BMME and HEOM results agree very well (see Fig.\ \ref{fig:charge_current_and_conductance_temperature_difference_induced} (a)).
For stronger coupling, the differences in the charge currents obtained with the HEOM and BMME methods (see Fig.\ \ref{fig:charge_current_and_conductance_temperature_difference_induced} (b)) are explained by the energy level broadening.
On the one hand, the charge current for small temperature differences is larger than predicted by the BMME because the broadening of the characteristic energy of transport processes enables the onset of these  processes earlier.
On the other hand, part of  the transport processes is shifted to higher energies, which become less effective when driven by a temperature difference. Accordingly, the charge current is reduced for large temperature differences by the broadening.

 \begin{figure}[!ht]
  \centering
  \footnotesize
   \includegraphics{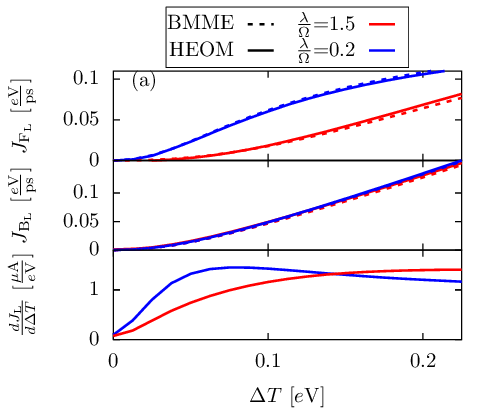}
   \includegraphics{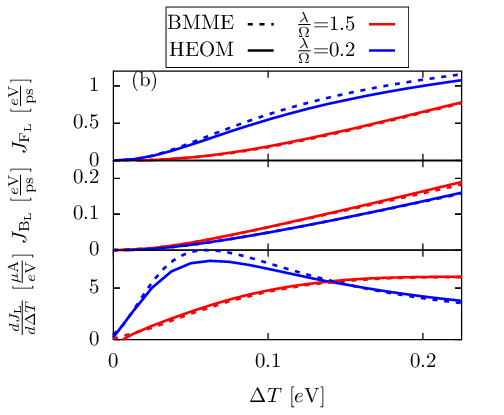}
  \caption{Heat currents out of the left contact for the bosonic and fermionic baths 
  and their added differential thermal conductivity as a function of temperature difference. In (a) and (b), the parameters are chosen according to parameter sets A and B, respectively. 
}
  \label{fig:heat_current_and_conductance_temperature_difference_induced}
 \end{figure}

Figure\ \ref{fig:heat_current_and_conductance_temperature_difference_induced} presents the heat current leaving the left electronic bath $J_{\text{F}_\text{L}}$, the heat current leaving the left bosonic bath $J_{\text{B}_\text{L}}$, and their added differential thermal conductivity $\frac{d J_\text{L}}{d \Delta T}=\frac{J_{\text{F}_\text{L}}+J_{\text{B}_\text{L}}}{d \Delta T}$ as a function of the temperature difference $\Delta T$.
In the weak system-environment coupling regime (see Fig.\ \ref{fig:heat_current_and_conductance_temperature_difference_induced} (a)), we observe an increase of the heat current out of the hot electronic bath in accordance with the behavior of the charge current. Due to Franck-Condon blockade of transport channels, the heat current for stronger electronic-vibrational interaction is reduced. 
In comparison to the charge current, the reduction is less pronounced as the processes with higher thresholds are realized by electrons that induce a vibrational energy transfer of multiple vibrational quanta.
The heat current out of the hot vibrational bath rises with increasing temperature difference without strong dependence on the electronic-vibrational coupling for a weak system-environment coupling.
 
Stronger system-environment couplings (see Fig.\ \ref{fig:heat_current_and_conductance_temperature_difference_induced} (b)) affect the electronic 
heat current analogous to the charge current.
As the cooling of the vibration by electrons gets more effective with increasing system-environment coupling,\cite{Haertle2018} the dependence on the electronic-vibrational coupling increases as well.

 \subsection{Computational aspects and numerical convergence}
 \label{sec:IA:Computational_aspects}
  
 We finally discuss the numerical aspects of the importance criterion and  the numerical convergence of the HEOM method.
  \begin{table}[!ht]
    \centering
    \begin{tabular}{c|rrrrr}
     \hline \hline
        m$\backslash$n & 0   &   1   &   2   & 3 & 4 \\
        \hline
        0  &  1 &  26 &  1\,001 &  19\,500 &  255\,775   \\ 
        1  &  4 & 104 &  4\,004 &  78\,000 & 1\,023\,100 \\ 
        2  & 10 & 260 & 10\,010 & 195\,000 & 2\,557\,750 \\ 
        3  & 20 & 520 & 20\,020 & 390\,000 & 5\,115\,500 \\ 
        4  & \;35 &\; 910 &\; 35\,035 &\; 682\,500 &\; 8\,952\,125 \\ \hline \hline
    \end{tabular}
    \caption{Total number of independent ADOs of type $\rho^{(m|n)}_{\bm{g}|\bm{h}}$ in the hierarchy for the bias voltage driven cases. One bosonic bath and 2 fermionic baths as well as 3 bosonic and 12 fermionic Pad\'e poles are used.}
    \label{tab:number_of_irreducible_ados_without_threshold_value}
 \end{table}
  For the computational treatment of the HEOM approach (see Eq.\ \eqref{eq:general_HEOM}), the number of explicitly considered ADOs is reduced, without loosing any information, by utilizing the symmetries of the ADOs within one tier,\cite{Jin2008,Haertle2013a,Kato2015} i.e.,
 \begin{subequations}  
  \begin{align}
   \rho^{(m|n)}_{{g_1 g_2{\cdot}{\cdot}{\cdot}g_m}|{h_1 {\cdot}{\cdot}{\cdot}h_n}} =& \rho^{(m|n)}_{{g_2 g_1 g_3{\cdot}{\cdot}{\cdot}g_m}|{h_1 {\cdot}{\cdot}{\cdot}h_n}}, \\
   \rho^{(m|n)}_{{g_1 {\cdot}{\cdot}{\cdot}g_m}|{h_1 h_2 {\cdot}{\cdot}{\cdot}h_n}} =& -\rho^{(m|n)}_{{g_2 g_1 g_3{\cdot}{\cdot}{\cdot}g_m}|{h_2 
h_1 h_3 {\cdot}{\cdot}{\cdot}h_n}}, \\
   \rho^{(m|n)}_{{g_1 {\cdot}{\cdot}{\cdot}g_m}|{h_1 {\cdot}{\cdot}{\cdot}h_n}} =& (-1)^m \rho^{(m|n)\dagger}_{{g_1 {\cdot}{\cdot}{\cdot}g_m}|{\bar h_n {\cdot}{\cdot}{\cdot}\bar h_1}}.
  \end{align}
 \end{subequations}  
  For an exemplary case, Table \ref{tab:number_of_irreducible_ados_without_threshold_value} shows the number of independent ADOs, which have to be 
considered in the different tiers of the HEOM. It demonstrates the strong 
increase of the number of ADOs with increasing order of the twofold hierarchy.
  Without the application of an importance criterion, the large number of 
ADOs in higher orders renders the numerical simulation intractable because all of them need to be saved in the memory.
 \begin{table}[b]
   \centering
   \begin{tabular}{c| r r r r r}
    \hline \hline 
    $o_\text{max}\backslash \theta$& \;$10^{-3}$   & \quad$10^{-5}$   & \qquad$10^{-7}$   & \qquad$10^{-9}$ &   0 \\ \hline
    2       & 39          & 330         &   1\,981      &  4\,957     &   
     15\,420 \\ 
    3       & 39          & 383         &   4\,057      & 21\,644     &   
  \,718\,480 \\ 
    4       & 39          & 383         &   4\,223      & 30\,570     &  \quad 19\,341\,210 \\
    \hline \hline
    \end{tabular}
    \caption{Number of ADOs which are considered within the HEOM for different truncation levels $\rho^{(m|n)}_{\bm{g}|\bm{h}}$ with $m,n\leq o_\text{max}$ and different importance threshold values $\theta$ for the parameters from set A (see Table \ref{tab:definition_parameter_sets}). 
    }
    \label{tab:number_of_ados_for_different_threshold_values_and_truncation_levels}
  \end{table}
  Table \ref{tab:number_of_ados_for_different_threshold_values_and_truncation_levels} shows that the importance criterion, Eq.\ \eqref{eq:importance_estimate}, reduces the number of independent ADOs which have to be considered very effectively for the two-fold fermionic-bosonic hierarchy. 
  
\begin{figure}[!ht]
   \centering
    \footnotesize
   \includegraphics{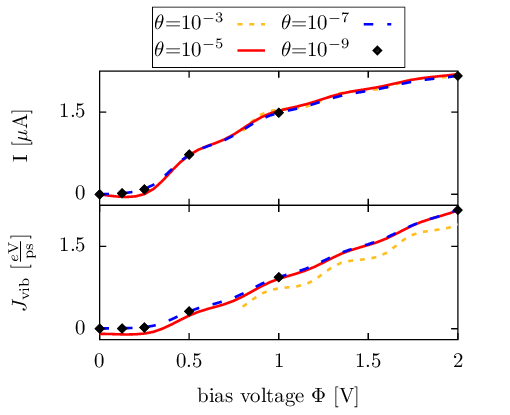}
   \caption{Convergence with respect to the threshold value $\theta$ 
for the charge and heat current as a function of bias voltage for parameter set B. Included are ADOs $\rho^{(m|n)}_{\bm{g}|\bm{h}}$ with $m,n\leq4$. }
   \label{fig:convergence_with_threshold_value}
  \end{figure}

  Next, we consider the consistency of this importance criterion.
  To this end, Fig.\ \ref{fig:convergence_with_threshold_value} shows exemplarily the effect of using the importance criterion on the convergence of charge or heat currents.
  Specifically, we depict the charge-current-voltage and the heat-current-voltage characteristics for different threshold values considered. 
  As the importance estimate for the ADOs within the fermionic part of the hierarchy is similar to the original definition by H\"artle \textit{et al.}\cite{Haertle2013a}, our observations in the bias voltage driven charge current are also similar.
  For large bias voltages, i.e., deep in the resonant transport regime, the results obtained with a high threshold value show little deviation from 
converged results.
  Approaching the nonresonant regime, $e\Phi < 2\overline{\varepsilon}_0$, these deviations become significant.
  For too large thresholds, we observe unphysical negative charge currents or even fail to obtain a stationary state result. The latter is caused by the occurrence of positive real parts in the eigenvalues of the propagation matrix of the system. The influence of these eigenvalues might be eliminated by the projection method suggested by Dunn \textit{et al.}\cite{Dunn2019}.

  The composition of ADOs in the heat current differs from that of the charge current, and thus, the convergence behavior with respect to the threshold value changes. We observe a deviation for large threshold values over the whole voltage bias window shown. This deviation decreases systematically by lowering the threshold value. As observed for the charge current, the convergence with respect to the threshold value is slower in the nonresonant regime.

  \begin{figure}[!ht]
   \centering
   \footnotesize
   \includegraphics{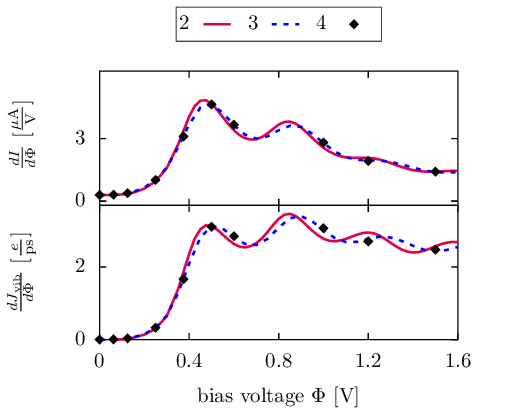}
   \caption{Convergence with respect to the truncation order for the 
charge and heat current as a function of bias voltage. Included are ADOs $\rho^{(m|n)}_{\bm{g}|\bm{h}}$ with $m,n\leq2,3,4$. The chosen parameters 
are $\overline{\varepsilon}_0=0.3\,\text{eV}, \frac{\lambda}{\Omega}=1.5,\Omega=0.2\,\text{eV}, T=0.025\,\text{eV}, \Lambda=0.0825\,\text{eV}, \Gamma=0.1\,\text{eV}$ with one vibrational bath and two fermionic baths.}
   \label{fig:convergence_with_order}
  \end{figure}
  Figure \ref{fig:convergence_with_order} illustrates the convergence of the HEOM results for a fixed threshold value of $\theta=10^{-9}$ with increasing truncation order $o_\text{max}$, which specifies the considered 
ADOs $\rho^{(m|n)}_{\bm{g}|\bm{h}}$ $m,n\leq o_\text{max}$.
  For the parameters considered, the results obtained for the different orders agree in the nonresonant regime, indicating convergence already for $o_\text{max}=2$. In the resonant regime, small shifts in the peak positions in the resonant regime are seen for lower $o_\text{max}$, which are more pronounced in the heat current than in the charge current. These are caused by frequency corrections introduced by higher-order effects of 
the vibrational bath, which require the inclusion of ADOs with at least $o_\text{max}=3$. 

  Finally, we comment on the computational effort required to solve the HEOM, especially the runtime, using as an example the data in Fig.\ \ref{fig:charge_current_and_conductance_bias_voltage_induced}. The data were obtained by the time-propagation of the HEOM into its stationary state using the Runge-Kutta-Fehlberg algorithm based on sparse matrix vector multiplication from the Intel \textsc{mkl} library. The runtimes for the different parameter regimes needed by a four-threaded execution on four CPUs of an Intel Xeon E6252 Gold processor are in the range of 5 h for parameter set A and 20 h for parameter set B. 
  The main reason for the different runtimes in the two parameters sets is the different system-environment coupling strengths. The number of ADOs, which have to be considered for converged results, increases strongly for larger coupling strengths. The effect of this ADO number increase on the runtime outweighs the effect of reduced propagation times caused by the enhanced damping of the internal system dynamics.      
  We also note that the computational effort does not depend significantly on the bias voltage and  the strength of the electronic-vibrational coupling.
  
%------------------------------
%Summary and outlook
%------------------------------

 \section{Conclusion}
 \label{sec:Conclusion}

In this work, we have formulated the HEOM method for open quantum systems 
involving multiple bosonic and fermionic environments. This extended approach allows the study of quantum transport through nanosystems induced by 
bias voltages and/or temperature differences in a numerically exact manner for both types of environments on equal footing. 
To facilitate efficient simulations with the extended HEOM method, we have also formulated an importance criterion for the elements of the hierarchical structure, which reduces the computational effort to a manageable level without compromising the ability to control the numerical convergence.

In order to demonstrate the performance of the extended HEOM method, we have applied it to a generic interacting model of electronic and vibrational charge and heat transport in a molecular junction. The results show the intricate interplay of electronic and vibrational degrees of freedom in 
this nonequilibrium transport scenario for both voltage and thermally driven transport processes. 

Applications of the extended HEOM method to more complex models and/or time-dependent setups are possible and promise to reveal further interesting physical effects in nanosystems. The approach may also provide new opportunities to study fundamental questions of quantum thermodynamics in nanosystems beyond linear response and weak coupling in a numerically exact way. \new{One example is fluctuations, which were studied recently using the HEOM method for bosonic or fermionic reservoirs separately.\cite{Cerrillo2016,Schinabeck2020}}{}
%which are contained in the full counting statistics and have been studied via a further extension of the HEOM for bosonic or fermionic reservoirs separately.\cite{Cerrillo2016,Schinabeck2020}}{}

\section*{Acknowledgments}

\new{We thank S. Wenderoth for fruitful discussions.}{}
This work was supported by a research grant of
the German Research Foundation (DFG).
Y.K. thanks the Alexander von Humboldt Foundation for the award of a Research Fellowship.
Furthermore, support by the state
of Baden-W\"urttemberg through bwHPC and the DFG
through Grant No. INST 40/575-1 FUGG (JUSTUS 2 cluster)
is gratefully acknowledged. 

\bibliography{./MyBib}
\end{document}